\documentclass[epj]{svjour}

\usepackage{graphicx}
\usepackage{amsmath, amssymb}
\usepackage{cite}
\usepackage{booktabs} 
\usepackage{tikz}
\usepackage{tikz-3dplot}
\usepackage{placeins}
\usepackage{tabularx}
\usepackage{hyperref}

\graphicspath{{figures/}} 

\usepackage{letltxmacro}
\LetLtxMacro{\originaleqref}{\eqref}
\renewcommand{\eqref}{Eq.~\originaleqref}

\usetikzlibrary{shapes, arrows}
\usetikzlibrary{decorations.pathreplacing}
\usetikzlibrary{positioning, calc}
\tikzstyle{fitted} = [
rectangle, minimum width=5cm, minimum height=1cm, text centered, draw=black, fill=red!30]
\tikzstyle{operations} = [rectangle, rounded corners, minimum width=2cm,text centered, draw=black, fill=red!30]
\tikzstyle{roundtext} = [rectangle, rounded corners, minimum width=2cm, minimum height=0.8cm, text centered, draw=black, fill=red!30]
\tikzstyle{n3py} = [rectangle, rounded corners, minimum width=3cm, minimum height=1cm, text centered, draw=black, fill=green!30]
\tikzstyle{myarrow} = [thick,->,>=stealth]
\tikzstyle{line} =[draw, -latex']
\tikzstyle{decision} = [diamond, draw, fill=red!20, text width=7.5em, text centered,  inner sep=0pt, minimum height=2em, aspect=4]
\tikzstyle{cloud} = [draw, ellipse,fill=green!20, minimum height=2em]
\tikzstyle{inout} = [rectangle, draw, fill=green!20, text width=9.5em, text centered, rounded corners, minimum height=2em, minimum width=10em]
\tikzstyle{block} = [rectangle, draw, fill=blue!20, text width=9.5em,
text centered, rounded corners, minimum height=2em,
minimum width=10em]

\begin{document}

\def\arraystretch{1.30}
\emergencystretch=1.5em

\title{A data-based parametrization of parton distribution functions}

\author{
    Stefano Carrazza\inst{1,2,3} \and
    Juan Cruz-Martinez\inst{1} \and
    Roy Stegeman\inst{1}
}
\institute{
    TIF Lab, Dipartimento di Fisica, Universit\`a degli Studi di Milano
    and INFN Sezione di Milano.
    \and CERN, Theoretical Physics Department, CH-1211 Geneva 23, Switzerland.
    \and Quantum Research Centre, Technology Innovation Institute, Abu Dhabi, UAE.
}
\date{Received: date / Revised version: date}
\abstract{
    Since the first determination of a structure function many decades ago, all
    methodologies used to determine structure functions or parton distribution functions (PDFs)
    have employed a common prefactor as part of the parametrization.
    The NNPDF collaboration pioneered the use of neural networks to overcome the inherent bias
    of constraining the space of solution with a fixed functional form while still keeping the same
    common prefactor as a preprocessing.
    Over the years various, increasingly sophisticated, techniques have been introduced to counter
    the effect of the prefactor on the PDF determination.
    In this paper we present a methodology to perform a data-based scaling of the Bjorken $x$ input parameter which facilitates the removal the prefactor, thereby
    significantly simplifying the methodology, without a loss of efficiency
    and finding good agreement with previous results.
} 
\authorrunning{S.C, J.C-M, R.S}
%

\maketitle

\section{Introduction}
Parton distribution functions (PDFs) provide a description of the
non-perturbative structure of
hadrons~\cite{Butterworth:2015oua,Kovarik:2019xvh,AbdulKhalek:2019mps}. An
accurate and precise description of PDFs is needed to make theoretical
predictions for precision physics at hadron colliders.
Since PDFs are non-perturbative, only lattice QCD allows for a numerical
approach to the problem~\cite{Constantinou:2020hdm}, otherwise it is not
possible to obtain the PDFs directly from first principles.
As a consequence, they are determined by performing fits to experimental data.
This requires the use of a
functional form of which parameters are to be fitted. Such a procedure can potentially lead to
a biased result, in some cases ultimately resulting in underestimated PDF
uncertainties~\cite{Ball:2008by}.

PDF fitting collaborations have developed different techniques to gauge and
reduce the impact of the different sources of bias and provide a correct
estimate of PDF uncertainties. Collaborations using the Hessian approach introduce the
notion of ``tolerance'', whereby the uncertainties are rescaled by a global
factor to obtain better agreement with the experimental data. It should be noted however, that tolerance is used to account for all sources of uncertainty, not just that corresponding to the parametrization but in particular it is also used to account for tensions between datasets and theoretical uncertainties. On the other hand, the NNPDF
collaboration addresses the problem of a biased functional form by
parametrizing the PDFs with neural networks. Neural networks can represent any
function, as is understood through the universal approximation
theorems~\cite{Cybenko1989}, thus avoiding biasing the PDF trough the choice of
the functional form.

In practical terms, however, the training of the parameters of a neural network
occurs in a finite number of steps and some assumptions are to be introduced to improve the efficiency of the training.
In the specific case of NNPDF, the input values are scaled according to their expected
distribution and the neural network is multiplied by a prefactor designed to speed up the training.
Great care has been put into ensuring that such preprocessing does not introduce a bias
in the PDF determination~\cite{Forte:2002fg} but their effect needs to be reevaluated
with every change to the methodology or dataset~\cite{Ball:2008by, Ball:2009mk, Ball:2017nwa}.

In this paper we present a purely data-driven approach for the preprocessing of the input data.
We implement it as part of the latest NNPDF fitting framework~\cite{Carrazza:2019mzf}
and analyze the resulting PDFs.
In both the data region and the extrapolation regions we obtain results
compatible with those of NNPDF4.0~\cite{Ball:2021leu}, with similar rates of
convergence.
This preprocessing methodology brings two important advantages,
firstly it facilitates the use of the NNPDF open source framework~\cite{NNPDF:2021uiq}
by external users by automatizing steps that until now required human intervention and,
secondly, it validates the NNPDF4.0 determination by removing two possible sources of bias or inefficiencies
without a significant change of the results.

The first possible source of bias is due to the input layer: in all previous releases of
NNPDF~(see e.g. Refs.~\cite{Ball:2014uwa,Ball:2017nwa,Ball:2021leu} for the
most recent releases) the input layer takes the momentum fraction $x$ and
splits it into a tuple $(x,\log x)$ which is provided as input to the dense
layers of the neural network.
As will be discussed in more detail in Sect.~\ref{sec:inputscaling}
the non-trivial interplay between the stopping algorithm and the
rates of convergence of two different input scales can lead to overfitted or underfitted
scenarios, putting an extra burden on the hyperoptimization algorithm.

The second potential source of bias is the prefactor $x^{(1-\alpha_{a})}(1-x)^{\beta_{a}}$ used by
all PDF fitting groups as part of their parametrization of the PDFs.
For most groups, such as MSHT~\cite{Bailey:2020ooq},
CTEQ~\cite{Hou:2019efy}, and ABMP~\cite{Alekhin:2017kpj}, this prefactor is
used as part of the functional form of the fitted PDFs (and thus its parameters are also fitted).
In the case of NNPDF instead the exponential parameters, $\alpha_{a}$ and $\beta_{a}$, are chosen randomly
in a per-replica basis from a pre-established range~\cite{Ball:2009mk, Ball:2014uwa}.
The random selection of exponents was introduced due to the observation in Ref.~\cite{Ball:2008by} that uncertainties
could be underestimated if the exponential parameters were to be fixed.
While as a result of this mitigation strategy the prefactor becomes just a tool
to speed-up convergence, it has a negative impact on aspects of the methodology
such as the scan of hyperparameters.
In Sect.~\ref{sec:nopreproc} we will discuss how, as a consequence of the improvements
of NNPDF4.0, the prefactor can be removed from the methodology
and thus also the mitigation strategies related to it.

The methodology we present is a feature scaling of the input
$x$, as is standard practice in the
machine learning community, leading to a simplification of the neural network architecture.
We show that PDFs resulting from these changes are faithful both in the data and the
extrapolation regions, and that they are compatible with the current generation
of NNPDF fits.

The paper is structured as follows. In Sect.~\ref{sec:nnpdfmeth} we will
highlight in more detail the parametrization choices made by PDF fitting
collaborations. In this section we will in particular focus on the NNPDF
methodology, as the main purpose of this paper is to present the feasibility
of the approach by applying it to the NNPDF framework. Then, in
Sect.~\ref{sec:inputscaling} we will discuss the new input scaling, and in
Sect.~\ref{sec:nopreproc} we will discuss how this change to the input scaling allows for the removal of
the prefactor. Finally in Sect.~\ref{sec:results} we validate the accuracy
and faithfulness of the PDFs by performing various tests
which fully validate the presented methodology.

\section{Parametrization of PDFs}
\label{sec:nnpdfmeth}

All of the most used PDF sets are parametrized at
some input scale $Q_0$ by a function of the form
\begin{equation}
    xf_{a}\left(x, Q_{0}\right)
    =A_{a} x^{(1-\alpha_{a})}(1-x)^{\beta_{a}}\mathcal{P}_{a}(y(x)),
    \label{eq:general_pdf_parametrization}
\end{equation}
where the indices $a$ correspond to the type of parton, and $\mathcal{P}_a$ is
a functional form that is different between PDF fitting groups, with an input
$y(x)$ that differs between PDF sets. In
particular,
for the most recent PDF sets released by the MSHT~\cite{Bailey:2020ooq},
CTEQ~\cite{Hou:2019efy}, and ABMP~\cite{Alekhin:2017kpj} collaborations
$\mathcal{P}_a$ represents a polynomial per flavor, while for the latest PDF set
released by the NNPDF collaboration~\cite{Ball:2021leu} $\mathcal{P}_a$ is
represented by a single neural network with one output node per flavor.

The PDFs are kinematically constrained by
\begin{equation}
    f_a(x=1,Q)=0,
    \label{eq:f(x=0)=1}
\end{equation}
which is enforced through the $(1-x)^{\beta_{a}}$ component in
\eqref{eq:general_pdf_parametrization}. The motivation to choose this
functional form stems from the constituent counting
rules~\cite{Brodsky:1973kr}. In the fitting methodologies this
component not only ensures that the condition of \eqref{eq:f(x=0)=1} is
satisfied, but it partially controls the large-$x$ extrapolation region where data is
unavailable.
The small-$x$ behavior instead is controlled by the prefactor $x^{(1-\alpha_{a})}$.
The introduction of this factor was inspired by Regge theory~\cite{Abarbanel:1969eh}.
While enforcing this behavior implies a methodological
bias~\cite{Forte:2020yip},
studies on the extrapolation behavior of PDF determinations have found
a qualitative agreement with the expected values~\cite{Ball:2016spl}.

Despite its generalized use, a fixed functional form as prefactor introduces
two issues in the determination of PDFs.
First, the PDFs are only based on data in the domain $10^{-5} \lesssim x \lesssim 0.75$
while PDF grids are delivered in the domain $10^{-9} \leq x \leq 1$ and as a result there is a potential
bias in the extrapolation regions.
Secondly, even if we assume that outside the data region the PDFs are
described by the given functional form,
it is not clear at which scale $Q^2$ it should hold and
it is not preserved under $Q^2$ transformation.

In this paper we will focus on the implications of this parametrization in the
context of the NNPDF methodology, thus where the PDFs are parametrized with a
neural network of which the output nodes correspond to a linear combination of PDF
flavors at an energy scale $Q_0$. The neural network consists of
fully-connected layers where the depth of the network and the number of nodes
in each layer, as well as the type of activation function, are determined
through a hyperoptimization procedure~\cite{Carrazza:2019mzf}.

The NNPDF model then can be written as
\begin{equation}
    \label{eq:oldmodel}
    xf_{a}\left(x, Q_{0}\right)
    =A_{a} x^{(1-\alpha_{a})}(1-x)^{\beta_{a}} \mathrm{NN}_{a}(x),
\end{equation}
where we recognize \eqref{eq:general_pdf_parametrization} with the general
function $\mathcal{P}_{a}$ replaced by a neural network $\mathrm{NN}_{a}$.
In the standard PDF fit the output flavors $a$ correspond the so-called
evolution basis, a linear combinations of PDF flavors that are eigenstates of
the $Q^2$ evolution~\cite{Ball:2021leu}, however, in Sect.~8.4 of Ref.~\cite{Ball:2021leu} it has been shown that different choices of the parametrization basis give good agreement within PDF uncertainties.
The prefactor $A_{a}$ is a normalization constant to
ensure that the momentum and valence sum rules are satisfied.

The effect of the exponents $\alpha_{a}$ and $\beta_{a}$ as a source of bias~\cite{Ball:2008by}
is mitigated in the NNPDF methodology by randomly sampling them from a uniform distribution,
the boundaries of which are determined through an iterative procedure~\cite{Ball:2009mk, Ball:2014uwa}.

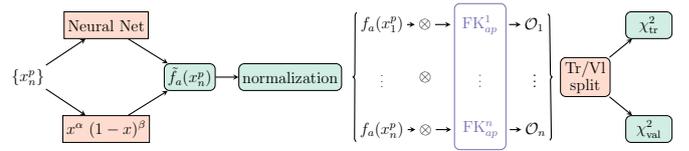
\begin{figure}[ht]
    \centering
    \resizebox{1.0\columnwidth}{!}{%
        \begin{tikzpicture}[node distance = 1.0cm]\scriptsize
            \definecolor{vp1}{RGB}{102,194,165}
            \definecolor{vp2}{RGB}{252,141,98}
            \definecolor{vp3}{RGB}{117,112,179}
            \node (xinput) {\large $\{x_{n}^{p}\}$};

            \coordinate [right = 1.5cm of xinput] (NNghost) {};
            \node[fitted, fill=vp2!30, above = 1.0cm of NNghost, minimum width=1.7cm, minimum height=0.7cm]
                (pdf) {\large Neural Net};
            \node[fitted, fill=vp2!30, below = 1.0cm of NNghost, minimum width=1.7cm, minimum height=0.7cm]
                (preproc) {\large $x^{\alpha}$ $(1-x)^{\beta}$};

            \node[operations, fill=vp1!30, minimum width=1.2cm, minimum height=0.7cm, right = 1.5cm of NNghost]
                (fitbasis) {\large $\tilde{f}_{a}(x^{p}_n)$};
            \node[operations, fill=vp1!30, minimum width=1.2cm, minimum height=0.7cm, right = 0.6cm of fitbasis]
                (normalizer) {\large normalization};

            \node[right = 0.9cm of normalizer] (pdfdots) {\vdots};
            \node[above = 0.7cm of pdfdots]
                (pdf1) {\large ${f}_{a}(x^{p}_1)$};
            \node[below = 0.7cm of pdfdots]
                (pdfn) {\large ${f}_{a}(x^{p}_n)$};

            \node[right = 0.2cm of pdf1] (conv1) {\large $\otimes$};
            \node[right = 0.2cm of pdfn] (convn) {\large $\otimes$};
            \node at ($(conv1)!0.5!(convn)$) (convdots) {\large $\otimes$};

            \node[vp3, right = 0.6cm of conv1] (f1) {\large FK$_{ap}^{1}$};
            \node[vp3, right = 0.6cm of convn] (fn) {\large FK$_{ap}^{n}$};
            \node[vp3] at ($(f1)!0.5!(fn)$) (fd) {\large \vdots};
            \draw[draw=vp3, rounded corners] ($(f1.north west)+(-0.1, 0.2)$) rectangle ($(fn.south east)+(0.1,-0.2)$);

            \node[right = 0.5 cm of f1] (o1) {\large $\mathcal{O}_{1}$};
            \node[right = 0.5 cm of fn] (on) {\large $\mathcal{O}_{n}$};
            \node at ($(o1)!0.5!(on)$) (od) {\large \vdots};

            \node[operations, fill=vp2!30, right = 0.5cm of od, minimum width = 1.2cm, text width=1.1cm, minimum height=0.7cm]
                (trvl) {\large Tr/Vl split};
            \coordinate [right = 1.0cm of trvl] (ending) {};
            \path let \p1 = (ending), \p2 = (pdf)
                in node at (\x1,\y2) [n3py, fill=vp1!30, minimum width = 1.2cm, minimum height=0.7cm] (tr) {\large $\chi^{2}_\text{tr}$};
            \path let \p1 = (ending), \p2 = (preproc)
                in node at (\x1,\y2) [n3py, fill=vp1!30, minimum width = 1.2cm, minimum height=0.7cm] (vl) {\large $\chi^{2}_\text{val}$};

            \draw[myarrow] (xinput) -- (pdf);
            \draw[myarrow] (xinput) -- (preproc);
            \draw[myarrow] (pdf) -- (fitbasis);
            \draw[myarrow] (preproc) -- (fitbasis);
            \draw[myarrow] (fitbasis) -- (normalizer);

            \draw[myarrow] (pdf1) -- (conv1);
            \draw[myarrow] (pdfn) -- (convn);
            \draw[myarrow] (conv1) -- ($(f1.west)-(0.2,0.0)$) ;
            \draw[myarrow] (convn) -- ($(fn.west)-(0.2,0.0)$) ;
            \draw[myarrow] ($(f1.east)+(0.2,0.0)$) -- (o1);
            \draw[myarrow] ($(fn.east)+(0.2,0.0)$) -- (on);

            \draw[myarrow] (trvl) -- (tr);
            \draw[myarrow] (trvl) -- (vl);

            \draw[decorate, decoration={brace}, thick] (pdfn.south west) -- (pdf1.north west);
            \draw[decorate, decoration={brace},thick] (o1.north east) -- (on.south east);
        \end{tikzpicture}
    }
    \caption{Diagrammatic representation of the calculation of the $\chi^{2}$
        in the NNPDF fitting framework as a function of the values of $\{x_n^{p}\}$
        for the different datasets.
    Each block indicates an independent component.}
    \label{fig:n3fit}
\end{figure}

The diagram in Fig.~\ref{fig:n3fit} shows how, within the NNPDF fitting
framework, the figure of merit ${\chi^2}$ is evaluated for an input grid
$\{x_n^{p}\}$, and how the prefactor and neural network are combined
to produce unnormalized PDFs $\tilde{f}_{a}(x^{p}_n)$, which are then
normalized by enforcing the momentum and valence sum rules~\cite{Ball:2021leu}.
The output of the parametrization then corresponds to normalized PDFs at an input scale $Q_0$.

The first layer of the network maps the input node $(x)$ onto the pair $(x,\log
x)$. The choice for this splitting of the input results from the
observation that typically PDFs show logarithmic behavior at small-$x$
($x\lesssim 0.01$) and linear behavior at large-$x$ ($x \gtrsim 0.01$)~\cite
{Forte:2002fg}, and together with the prefactor it ensures
convergence of the optimization algorithm in the small-$x$ region.

The PDFs themselves do not allow for a direct comparison to data, instead we
convolute the PDFs with partonic cross-sections to calculate a theoretical
prediction of physical observables corresponding to the experimental
measurements. These partonic cross-sections, along with the QCD evolution
equations, are encoded in \texttt{FastKernel} (FK) tables~\cite{Bertone:2013vaa,Bertone:2016lga},
allowing for an efficient computation of the relevant observables. For hadronic
observables the corresponding calculation is
\begin{equation}
    \mathcal{O}_{n}=\mathrm{FK}_{abpq}^{n} f_{a}(x_n^{p},Q_0)f_{b}(x_{n}^q,Q_0),
    \label{eq:prediction_dy}
\end{equation}
while for deep inelastic scattering observables it reduces to
\begin{equation}
    \mathcal{O}_{n}=\mathrm{FK}_{a p}^{n} f_{a}(x_n^p,Q_0),
    \label{eq:prediction_dis}
\end{equation}
where $n$ labels the experimental datapoints, $a$ and $b$ the PDF
flavors, and $p$ and $q$ the point in the corresponding $x$-grid.

After calculating the observables as predicted by the PDFs, the figure of merit
minimized during a fit is defined as
\begin{equation}
    \chi^{2}=\sum_{i, j}^{N_{\text {dat }}}(D-P)_{i} C_{ij}^{-1}(D-P)_{j},
    \label{eq:chi2=}
\end{equation}
with
\begin{equation}
    C_{ij} = C_{ij}^{\rm (exp)}.
    \label{eq:cov=exp}
\end{equation}
Here $D_i$ is the experimentally determined value of the $i$-th datapoint,
$P_i$ is
the theoretical prediction of the $i$-th datapoint calculated using
\eqref{eq:prediction_dy} or \eqref{eq:prediction_dis}, and
$C_{ij}^{\rm(exp)}$ is the covariance matrix of the experimental dataset.
For the optimization of the neural network the covariance matrix is treated
following the $t_0$ prescription of Ref.~\cite{Ball:2012wy}.

The final predictions (and experimental data) are then split into training and
validation sets such
that only the $\chi^{2}$ of the trained data is used for the minimization
while the validation $\chi^{2}$ is utilized for early stopping.

\section{A data-based scaling of the input $x$-grids}
\label{sec:inputscaling}

Often, in machine learning problems, the input data can be unbalanced or span several orders of magnitude.
Such is the case of PDF fitting, where the input is concentrated at small-$x$.

This can be a problem because, as we will explicitly show below, having input
features of different magnitudes introduces an artificial impact on the
importance of each feature within the network. This problem is exacerbated for
gradient descent based algorithms where the issue propagates to the learning
rate of the weights of the network. Thus, even if the algorithm is still able to
find the global minimum, the rate of convergence is not equal for all features.
In the case we are interested in (the NNPDF methodology with an early stopping
algorithm) this can lead to locally overfitted or underfitted results in
different regions of the kinematic domain. Ideally the fitting methodology
should result in a uniform rate of convergence across all input scales.

In short, the problem is that while the data spans multiple orders of magnitude,
the fitting methodology requires the inputs to be of the same length scale.
Below we discuss the impact of the input scaling on the PDFs, and provide a
methodology that takes an arbitrary input grid and scales it such that the
optimizer always has a good resolution across the entire input grid.

At this point one may note that the input $x$-grids to the neural network are
the grids defined in the FK-tables as shown in
Eqs.~(\ref{eq:prediction_dy},\ref{eq:prediction_dis}), which may differ from the
$x$-values of the corresponding experimental datasets. While this is true, from
the perspective of the fitting methodology, the grid choice is arbitrary and
thus the problem remains. 

In NNPDF fits, the input variable is mapped to $(x,\log x)$ in the first layer of the neural
network which facilitates the methodology in learning features of the PDF that scale
either linearly or logarithmically in $x$.
This splitting was first introduced in Ref.~\cite{Forte:2002fg} and was motivated by the
expectation that they are the variables upon which the structure functions
$F_2$ depend. It was noted that, by merit of the neural network-based
parametrization, the choice of input scales could affect the rate of
convergence but not the final result.

\begin{figure}[t]
    \includegraphics[width=1\columnwidth]{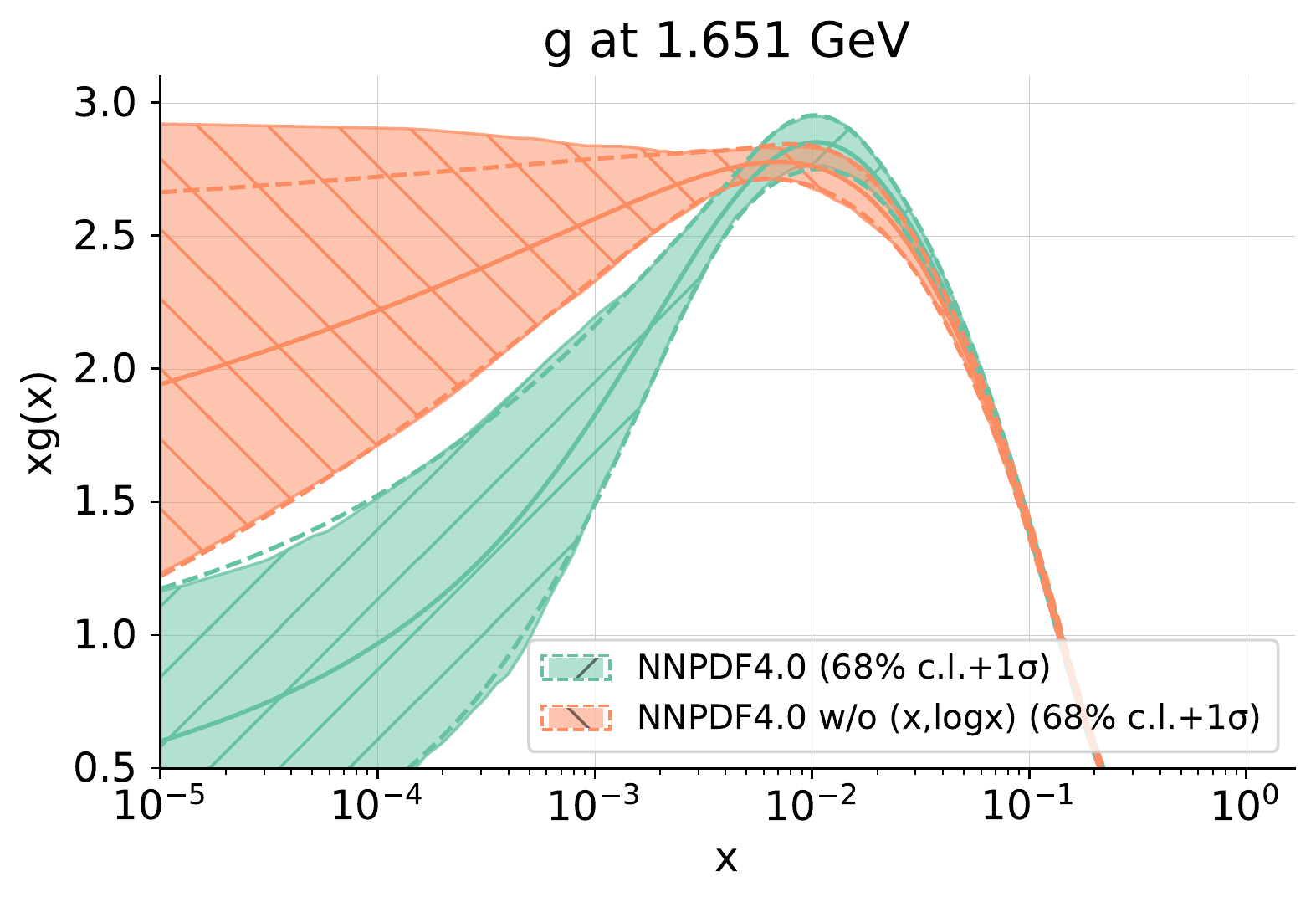}
    \caption{Comparison between the gluon PDF generated with the standard
        NNPDF4.0 methodology (green) and our modification in which we have removed
        the splitting layer of $x$ to $(x, \log x)$ (orange). While we observe good
        compatibility between both PDFs in the large-$x$ region, as we enter in the
        small-$x$  region our modified PDF saturates. This is evident also in the $\chi^{2}$
    of the modified fit which was blocked at $\chi^{2}=1.20$ while NNPDF4.0 is able to get it down to $\chi^{2}=1.16$.}
    \label{fig:without_x_logx}
\end{figure}

However, it can be seen in Fig.~\ref{fig:without_x_logx} that the $(x, \log x)$ split
can have an effect on the shape of the PDFs.
In Fig.~\ref{fig:without_x_logx} we compare the gluon PDF of the NNPDF4.0 fit
to a PDF generated using the same data, theory and methodological settings,
but with the $(x,\log x)$ input scaling replaced with only an $(x)$
input\footnote{This plot, as well as the other PDF plots
    in this paper, have been produced with the
\texttt{validphys} package~\cite{zahari_kassabov_2019_2571601, NNPDF:2021uiq}.}.

While the NNPDF4.0 methodology was sensitive to the small-$x$ region (where the logarithmic behavior is expected)
when we remove $(\log x)$ from the input we can observe a hint of saturation in said region.
Despite the fact that the $(x)$ and $(\log x)$ variables contain the same information
the split has a noticeable effect on the fit,
in this case speeding up the minimization process in the small-$x$ region.

\begin{figure}[ht]
    \includegraphics[width=1\columnwidth]{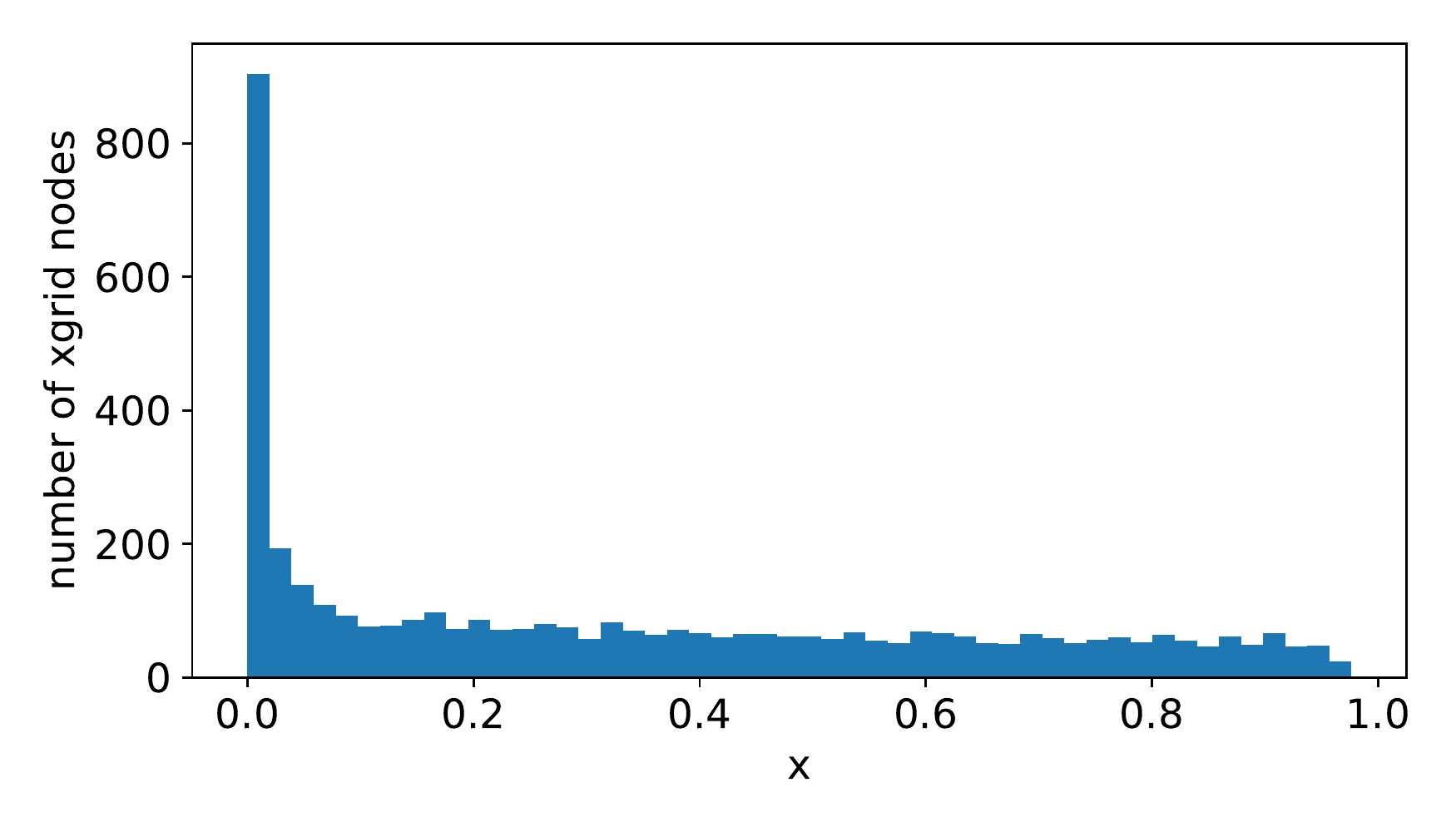}\\
    \includegraphics[width=1\columnwidth]{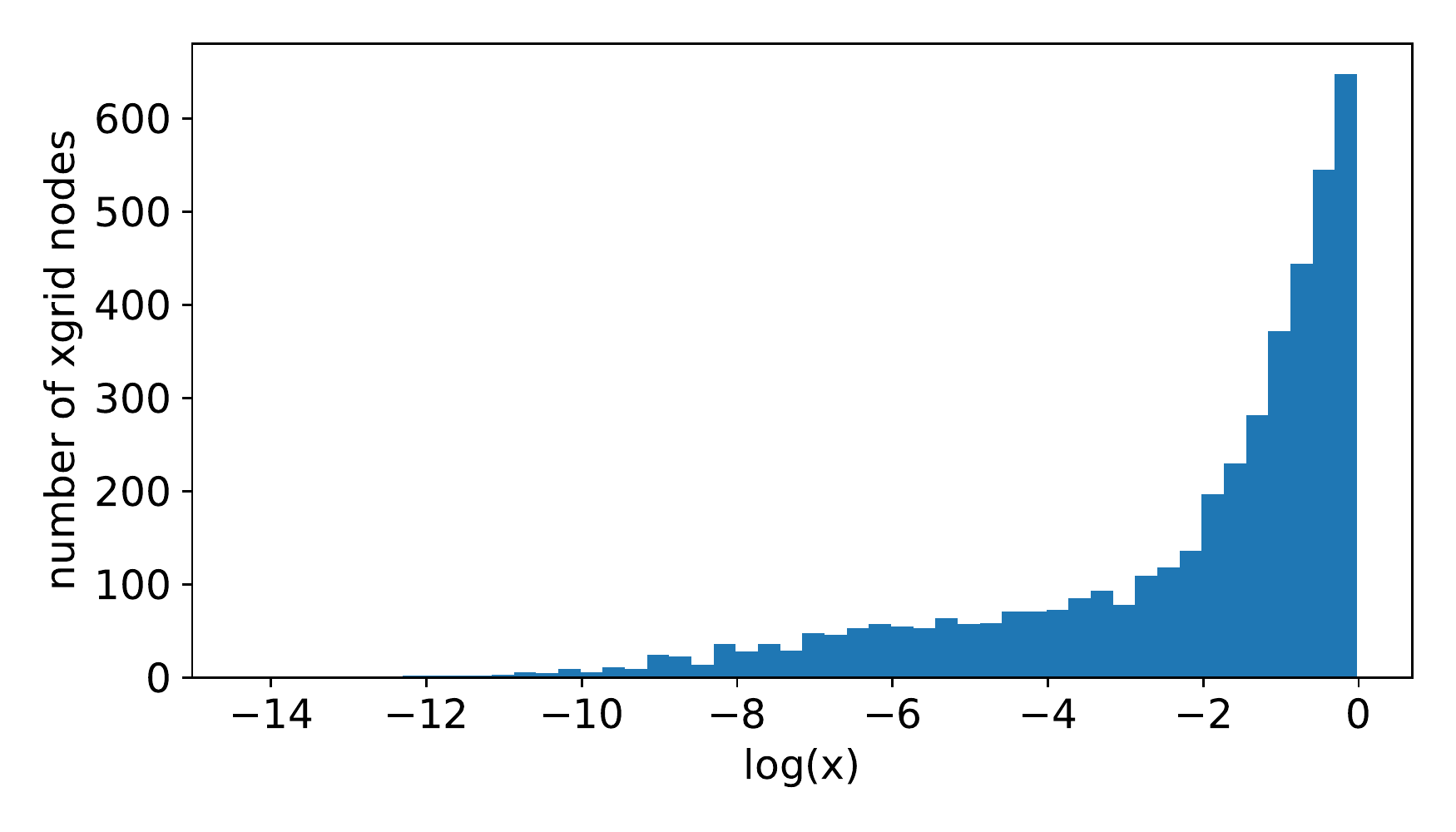}\\
    \includegraphics[width=1\columnwidth]{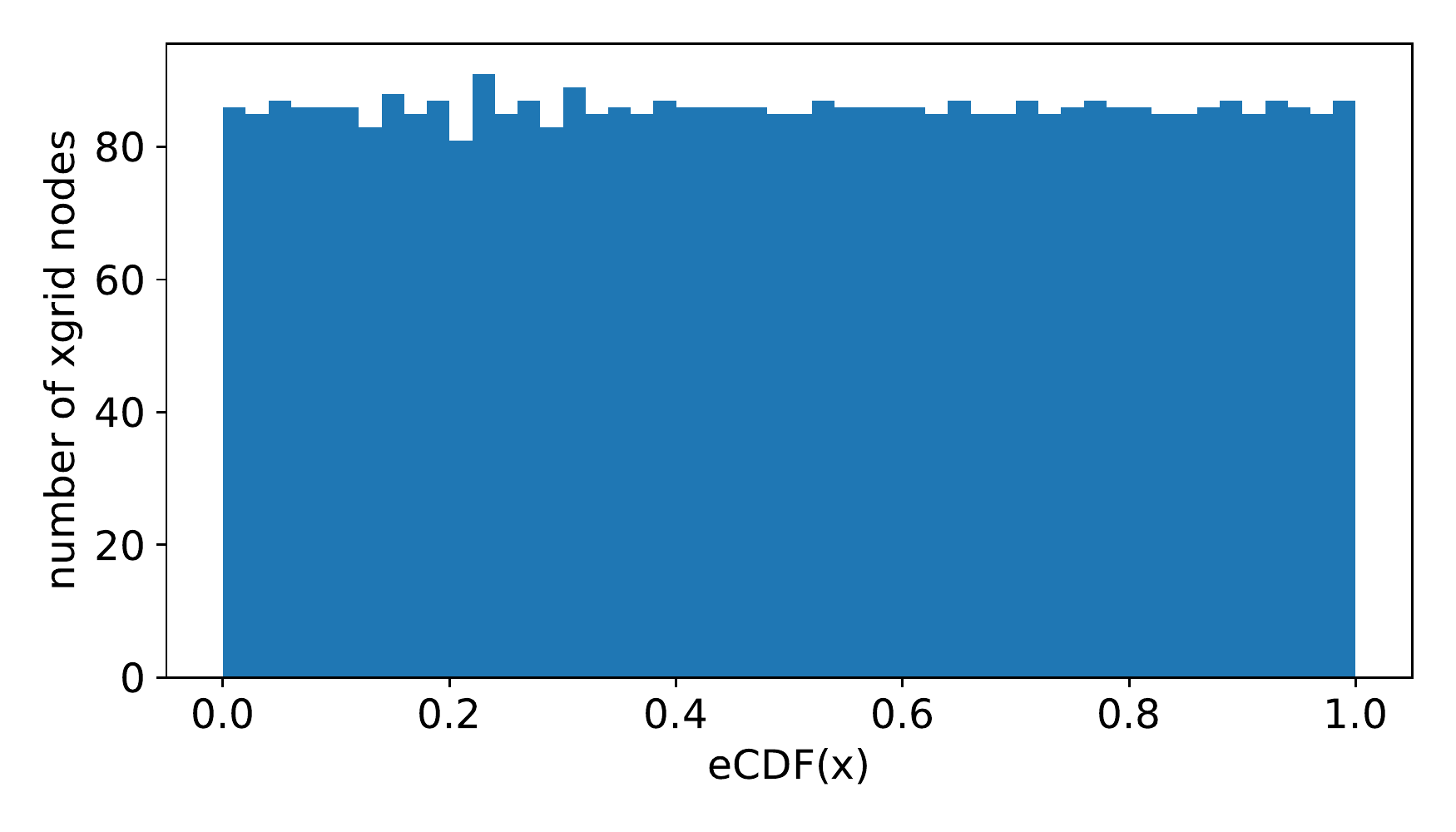}
    \caption{Histograms showing the distribution of the unscaled $x$ points in the
        FK-table $x$-grids (top), as well as the distribution of the input points after scaling with $\log x$ (middle) and eCDF (bottom).
    }
    \label{fig:ecdf_hist}
\end{figure}

In order for the optimization algorithm to be able to easily learn features across
many orders of magnitude we can perform a feature scaling of the training input $x$ such
that the distances between all points are of the same order of magnitude. In
particular, we can consider mapping the combined training input $x$-grid from
the FK-tables of all datasets as discussed in Sect.~\ref{sec:nnpdfmeth} to an
empirical cumulative distribution function (eCDF) of itself. The eCDF is
defined as a step function that starts at 0 and increases by $1/N_x$ at each
point of the input $x$-grids, with $N_x$ the total number of nodes in the
$x$-grids. If the $x$-grids of $n$ FK-tables share a common point in $x$, the
step-size corresponding to this point is instead $n/N_x$. This results in a
function whose value at any ${x}$ corresponds to the fraction of points
in the $x$-grids that are less than or equal to ${x}$. In other words, while
the $x$ values present in the FK table are not uniformly distributed on the
domain $0\leq x \leq 1$, applying the eCDF makes it that they are. A density
plot of the distribution of input points without scaling, logarithmically
scaled, and after applying the eCDF is shown in Fig.~\ref{fig:ecdf_hist}. This
figure also clearly shows that both inputs to the neural network as used in
NNPDF4.0~\cite{Ball:2021leu} have a high density of points on the same scale.

Applying the eCDF results in a distribution on the domain $0\leq x \leq
1$. However, for the results presented in this paper the eCDF transformation
is followed by a linear scaling, resulting in a total transformation of the
input $\hat{x} = 2 \cdot \mathrm{eCDF(x)}-1$, meaning that the input values to the neural
network are in the range $-1\leq x \leq 1$. This is done to ensure that the
input is symmetric around 0 which results
in improved convergence for many of the commonly used activation functions in
neural networks.

Since using the eCDF means that we apply a discrete scaling
only for values present in the input $x$-grids, we need to also add both an
interpolation and an extrapolation function to extract PDF values at values of
the momentum fraction that do not coincide with the input $x$-grids. Here it is
important to note that the PDFs are made publicly available through the
LHAPDF interface, and that they are correspondingly stored in the
LHAPDF grid format~\cite{Buckley:2014ana}. Because
LHAPDF grids are provided on the domain $10^{-9}\leq x \leq 1$, the problem of
extrapolation can be turned into an interpolation problem by including the
points $x=10^{-9}$ and $x=1$ in the input $x$-grid before determining the eCDF,
and defining a methodology for interpolation.

The simplest option for an interpolation function is a ``nearest neighbor'' mapping, whereby we map any input on the
continuous domain $0\leq x \leq 1$ to the nearest node in the $x$-grids of the
FK-table.

We can a nevertheless improve this simple mapping by instead using a continuous function.
A requirement of any such interpolation function is that it needs to be monotonically increasing.
However, if we determine the interpolation between each two points of the
FK-table $x$-grids the optimization algorithm will be agnostic to the existence
of this interpolation function as it is never probed. Ideally, in particular
for the evaluation of validation data of which the corresponding FK-tables were
not included when defining the eCDF scaling, we want the optimizer to probe the
interpolation functions such that it is able to learn its properties and as a
result provide a more accurate prediction in the interpolation region as well.

As such, the interpolation
functions are not defined between each neighboring pair of values in the input
$x$-grid, but rather we select $N_\mathrm{int}$ evenly distributed points
(after the eCDF transformation) between which to define interpolation
functions. Here $N_\mathrm{int}$ is a new hyperparameter, though not
necessarily one that needs to be free during hyperoptimization of the
methodology.
To obtain a monotonic interpolation function, we propose determining the interpolation
functions using cubic Hermite
splines~\cite{fritsch1980monotone}.

\begin{figure}[ht]
    \includegraphics[width=1\columnwidth]{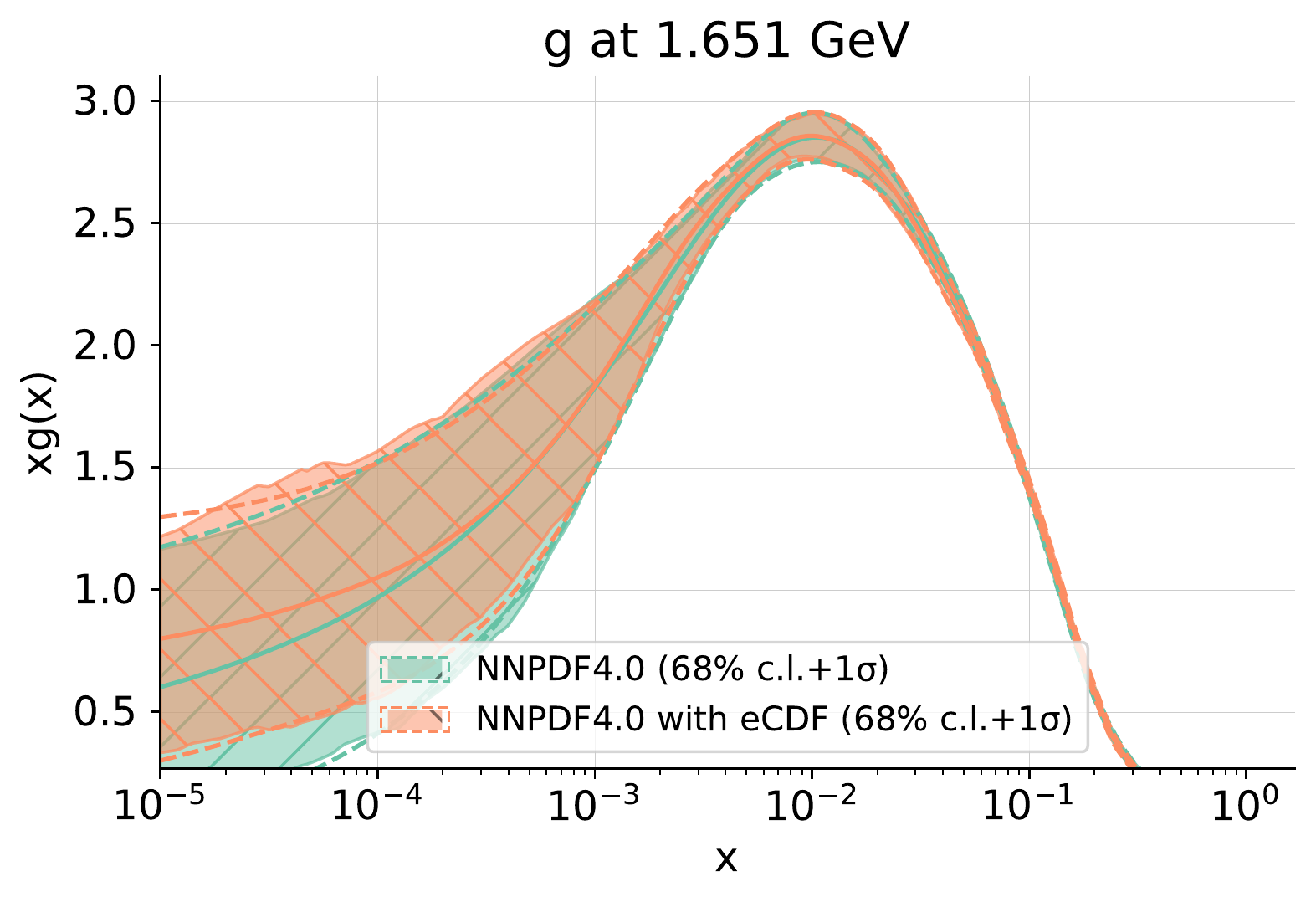}\\
    \includegraphics[width=1\columnwidth]{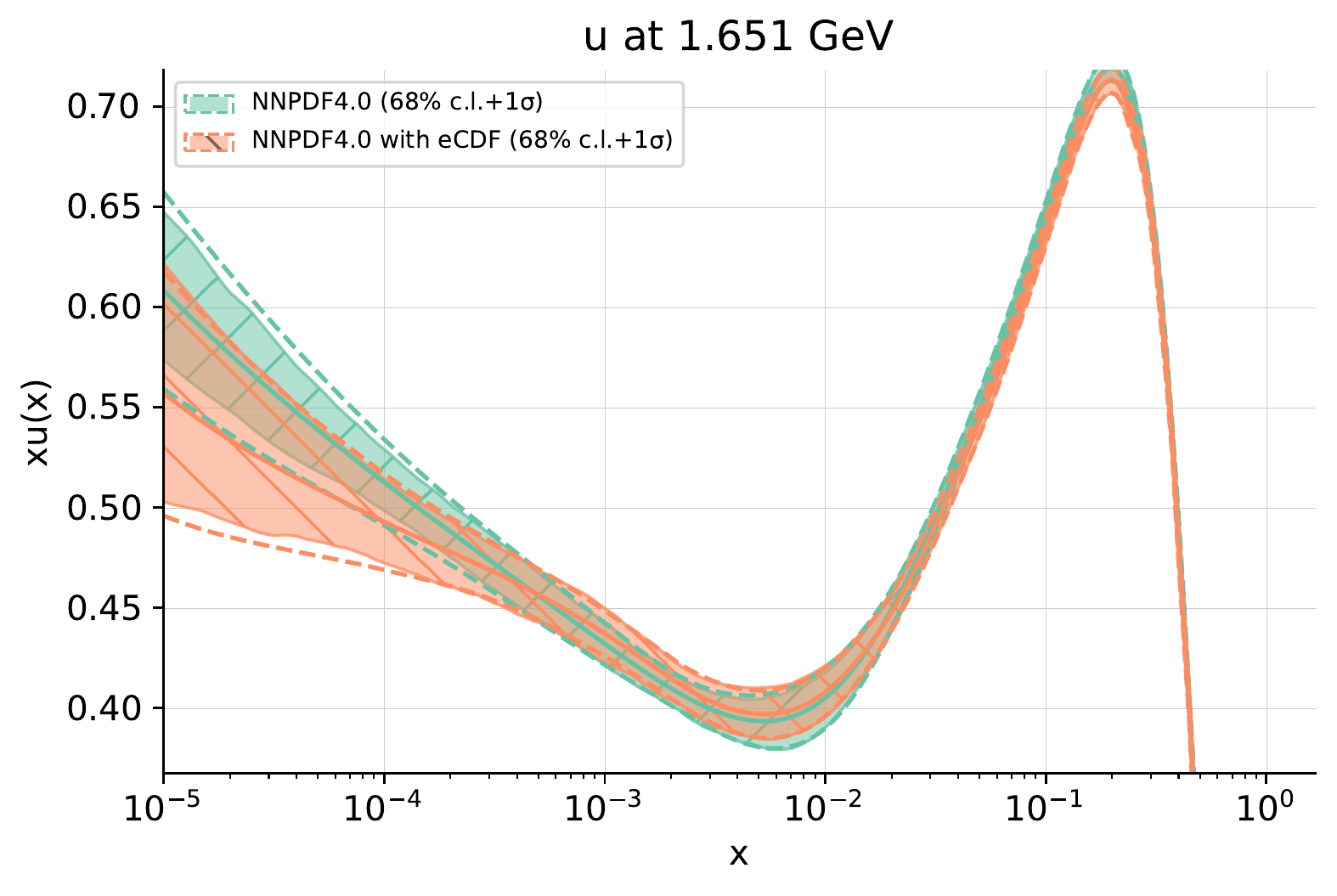}
    \caption{Comparison between the gluon and up PDFs determined using the NNPDF4.0 methodology (green)
    and a PDF determined using input scaling based on the eCDF (orange) with all other parameters the same.}
    \label{fig:pdf_inputscaling}
\end{figure}

By scaling the input as discussed in this section we remove any restrictions on
the PDF resulting from the input features, while simultaneously simplifying the
model architecture by getting rid of the mixing of two different orders of
magnitude in the first layer. In Fig.~\ref{fig:pdf_inputscaling} we compare the
gluon PDF generated using the NNPDF4.0 methodology, to a PDF generated using the same data and theory settings,
but with the $(x,\log x)$ input scaling replaced with the eCDF input scaling as
described above. This comparison of the gluon PDF is representative for all
flavors, and shows that the PDFs produced with this new scaling are in
agreement with those found using the $(x,\log x)$ input.
If the PDFs had not been in agreement that would have suggested that the PDFs
had a component that scaled neither linearly nor logarithmically
which the neural network was not able to accommodate.

\section{Removing the prefactor}
\label{sec:nopreproc}
In Sect.~\ref{sec:inputscaling} we discuss a new way of treating the input for
the PDF fitting by rescaling the input in a systematic way that depends only on
the fitted data itself.
This is a purely data-driven approach and thus free of sources of bias due to
the choice of functional form. As we will show here, the scaling of the input
grid in $x$ by using the eCDF will also allow us to remove the prefactor entirely.

In what follows we will discuss the consequence of removing the prefactor.
Specifically, by ``removing the
prefactor'', we understand a treatment which is equivalent to
setting $\alpha_a=1$ and $\beta_a=0$ in \eqref{eq:general_pdf_parametrization},
while enforcing the condition of \eqref{eq:f(x=0)=1}. As a result the PDF model
is written as
\begin{equation}
    xf_a(x,Q_0) = A_a \left[\mathrm{NN}_a(x) -\mathrm{NN}_a(1) \right].
\end{equation}
A similar model, without the model-agnostic input scaling, has previously been
applied to the study of fragmentation functions~\cite{Bertone:2017tyb}.
We will focus on the effects of the
change in the small-$x$ and large-$x$ extrapolation regions where the lack of
data makes the fit particularly prone to methodological biases.

In Sect.~\ref{sec:nnpdfmeth} it was mentioned that the motivation to include
the prefactor in NNPDF is to improve convergence during optimization
and that its effect as a source of bias in the extrapolation region
was mitigated by randomly sampling the exponents $\alpha_a$ and $\beta_b$.
However, while this makes the fit robust with respect to the exponents
it still comes with a cost:
it introduces an additional source of fluctuations between replicas
which can be undesirable in certain cases and limit progress around the
methodological development.

As a result of switching to optimizers based on stochastic gradient descent
(SGD) for NNPDF4.0, instead of using the genetic algorithm used for NNPDF3.1,
the average time to fit a replica has been reduced by an order of magnitude,
and stability has greatly improved~\cite{Carrazza:2019mzf,Ball:2021leu}.
These improvements of the optimization algorithm allow us to remove the
prefactor without a significant
change in computational costs
and therefore any possible benefit of the
prefactor in terms of convergence no longer outweighs its disadvantages.

As an example of where fluctuations between replicas as a result of the
randomized exponents of the prefactor can limit the development of the
methodology, one can consider the hyperoptimization procedure introduced in
Ref.~\cite{Carrazza:2019mzf} and further developed in Ref.~\cite{Ball:2021leu}.
The hyperoptimization algorithm employs out-of-sample testing through $K$-folds
cross-validation. In the current scenario an otherwise good hyperparameter setup
with poor exponents in the prefactor can return a worse figure of merit during
hyperparameter optimization than a relatively poorer hyperparameter setup with
very suitable exponents.
As a result many more hyperparameter combinations need to be tested to overcome
the statistical noise.
Removing the replica-by-replica random
sampling of the exponents removes this effect from hyperoptimization.

The uncertainties of the fit in the extrapolation region are closely related to
the ranges the prefactor exponents are sampled from. Removing them from the
parametrization also removes the random sampling, it is therefore important to
validate the obtained small-$x$ and large-$x$ uncertainties as will be done in
Sect.~\ref{sec:results}.

For brevity and clarity, we will from now on refer to the proposed methodology
without the prefactor and with the eCDF input scaling as the ``feature
scaling'' methodology.

\section{Results and validation}
\label{sec:results}

\subsection{Tuning the methodology: hyperoptimization}
After any significant change to the fitting methodology, it is important to
re-evaluate the choice of the hyperparameters of the model. The model parameters
obtained through the hyperoptimization procedure are given in Tab.~\ref{tab:setup}.
The details of the procedure are described in Sect.~3.3 of Ref.~\cite{Ball:2021leu}.
Note that the
selected activation function does not saturate asymptotically for large or small
values of $x$, preventing saturation in the extrapolation region. The choice of
activation function was however not fixed during the selection of
hyperparameters, this activation function has been selected by the
hyperoptimization algorithm among a selection of both saturating and
non-saturating activation functions.

\begin{table}[ht]
    \scriptsize
    \centering
    \renewcommand{\arraystretch}{1.4}
    \begin{tabularx}{\columnwidth}{XXXX}
        \toprule
        Parameter                                    &  Flavor basis             \\
        \midrule
        Architecture                                 &  1-59-49-48-42-8          \\
        Activation function                          &  $|x|\tanh (x)$           \\
        Initializer                                  &  \texttt{glorot\_normal}  \\
        Optimizer                                    &  \texttt{Nadam}           \\
        Clipnorm                                     &  1.5$\times 10^{-5}$      \\
        Learning rate                                &  4.3$\times 10^{-3}$      \\
        Maximum \# epochs                            &  19$\times 10^{3}$        \\
        Stopping patience                            &  24\% of max epochs       \\
        Initial positivity $\Lambda^{(\rm pos)}$     &  34                       \\
        Initial integrability $\Lambda^{\rm (int)}$  &  10                       \\
        $N_{\rm int}$                                &  40                       \\
        \bottomrule
    \end{tabularx}
    \vspace{0.2cm}
    \caption{The hyperparameter configuration selected using the $k$-folds
        hyperoptimization and used to perform the ``feature scaling'' fits presented
    in this paper.}
    \label{tab:setup}
\end{table}

\begin{figure}[ht]
    \includegraphics[width=0.48\textwidth]{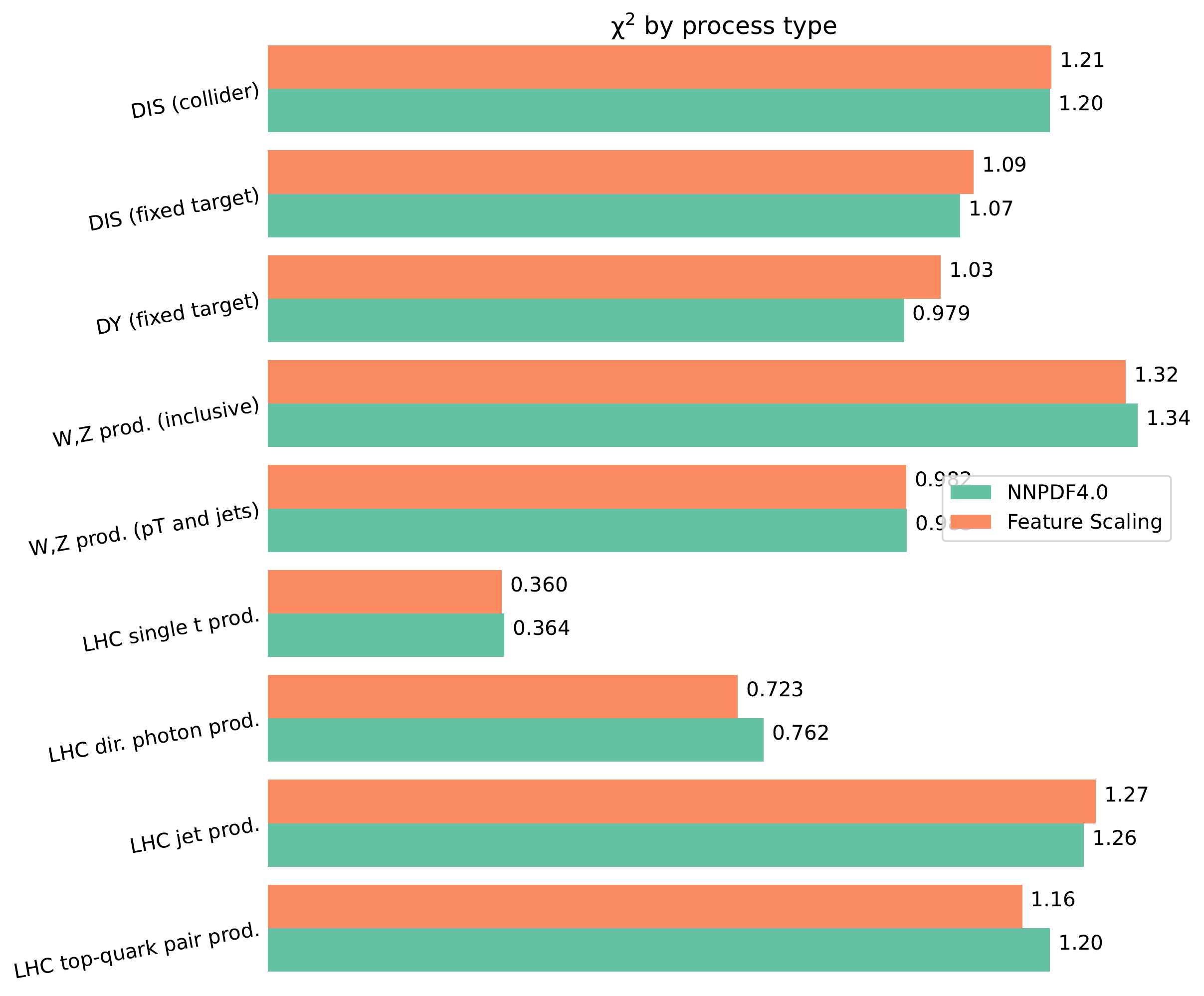}
    \caption{A comparison of the ${\chi^2}$ per process type between NNPDF4.0 (green) and feature scaling (orange), the total $\chi^2$ of feature scaling is 1.17 while that of NNPDF is 1.16.}
    \label{fig:chi2_feature_vs_nnpdf40}
\end{figure}

Having identified the best settings for the hyperparameters, we can analyze the
effect that changing the parametrization has on the PDFs and the predictions
made with them.
the ${\chi^2}$ values obtained with the updated methodology are shown in Fig.~\ref{fig:chi2_feature_vs_nnpdf40} where they are compared to those of NNPDF4.0.
From this it is clear that the feature scaling methodology provides a fit
to the data as good as NNPDF4.0.

In what follows we will study the implications of the
methodology in more detail, in many cases by comparing it to a PDF based on the
same experimental dataset and theory setting, but produced using the NNPDF4.0
methodology. Specifically, we will perform various tests to validate the PDFs
both in the extrapolation regions (see Sect.~\ref{sec:futuretest} and
Sect.~\ref{sec:largex}), as well as the data region (see
Sect.~\ref{sec:closuretest}). These test comprise the validation of the NNPDF4.0
methodology, and we will show that the performance of feature scaling is very
similar to that of NNPDF4.0.

\subsection{Validation of small-\texorpdfstring{$x$}{x} extrapolation}
\label{sec:futuretest}

To begin with, we need the PDFs to accurately describe the kinematic
domain from which the methodology has not seen data during training. If we
are able to determine the ${\chi^2}$ for this unseen data, that would
provide some insight into the generalization of our methodology in the
extrapolation region.

By definition, testing the accuracy in a region where there is no data to test
against is impossible.
Given that waiting for a future collider to become operational could take
decades, the next best thing we can do is to perform a fit to a ``historic''
dataset representing the knowledge available at an earlier point in time.
To this end we utilize the ``future test'' technique introduced in
Ref.~\cite{Cruz-Martinez:2021rgy}, and used to validate the small-$x$
extrapolation region of the NNPDF4.0 PDFs in Ref.~\cite{NNPDF:2021uiq}.
For consistency we keep the same datasets as presented in the original future
test paper (pre-HERA and pre-LHC).
In short, the test goes as follows: if the prediction from our methodology
is able to accommodate (within uncertainties) currently available data that
was not included in the fit, then the test is successful 
and we consider the generated uncertainties to be faithful.

Since the aim of doing a future test is to determine the ability of a
methodology for PDF determination to provide a generalized fit, we need to take
into account not only the uncertainty of the experimental data but also the
uncertainty of the PDF itself. This is done by redefining the covariance matrix
in \eqref{eq:chi2=} as
\begin{equation}
    C_{ij} = C_{ij}^{\rm (exp)}  + C_{ij}^{\rm (pdf)},
    \label{eq:cov=exp+pdf}
\end{equation}
where $C_{ij}^{\rm (pdf)}$ corresponds to the covariance matrix of the observables calculated from PDF predictions. Specifically,
$C_{ij}^{\rm (pdf)}$ is defined as
\begin{equation}
    {C}_{ij}^{\rm (pdf)} = \langle \mathcal{F}_i\mathcal{F}_j  \rangle_{\rm rep} - \langle \mathcal{F}_i  \rangle_{\rm rep}\langle \mathcal{F}_j  \rangle_{\rm rep},
\end{equation}
where $\mathcal{F}^{k}_i$ is the prediction of the $i$-th datapoint using the
$k$-th PDF replica with the average defined over replicas.

\begin{figure}[ht]
    \includegraphics[width=0.99\columnwidth]{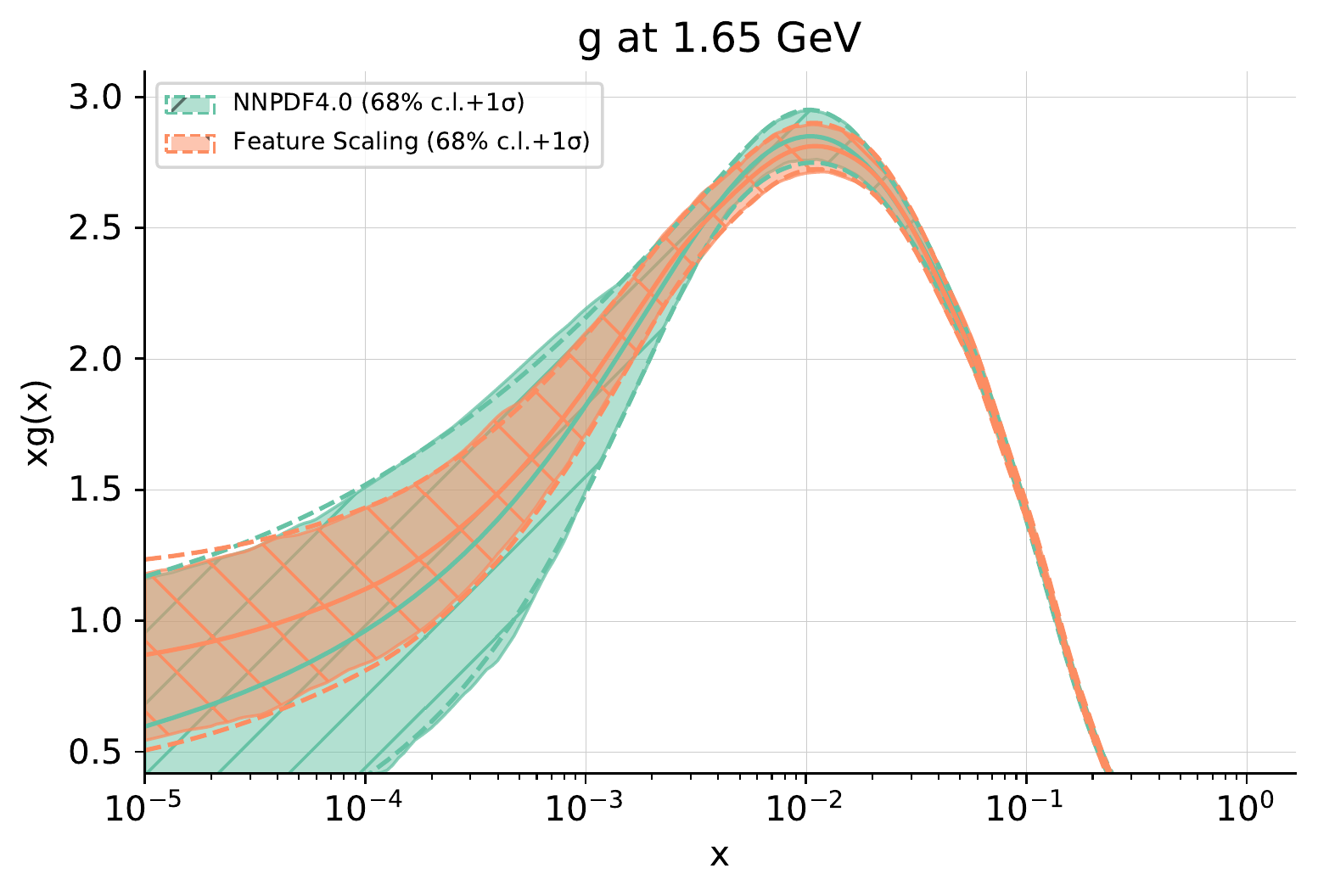}
    \includegraphics[width=0.99\columnwidth]{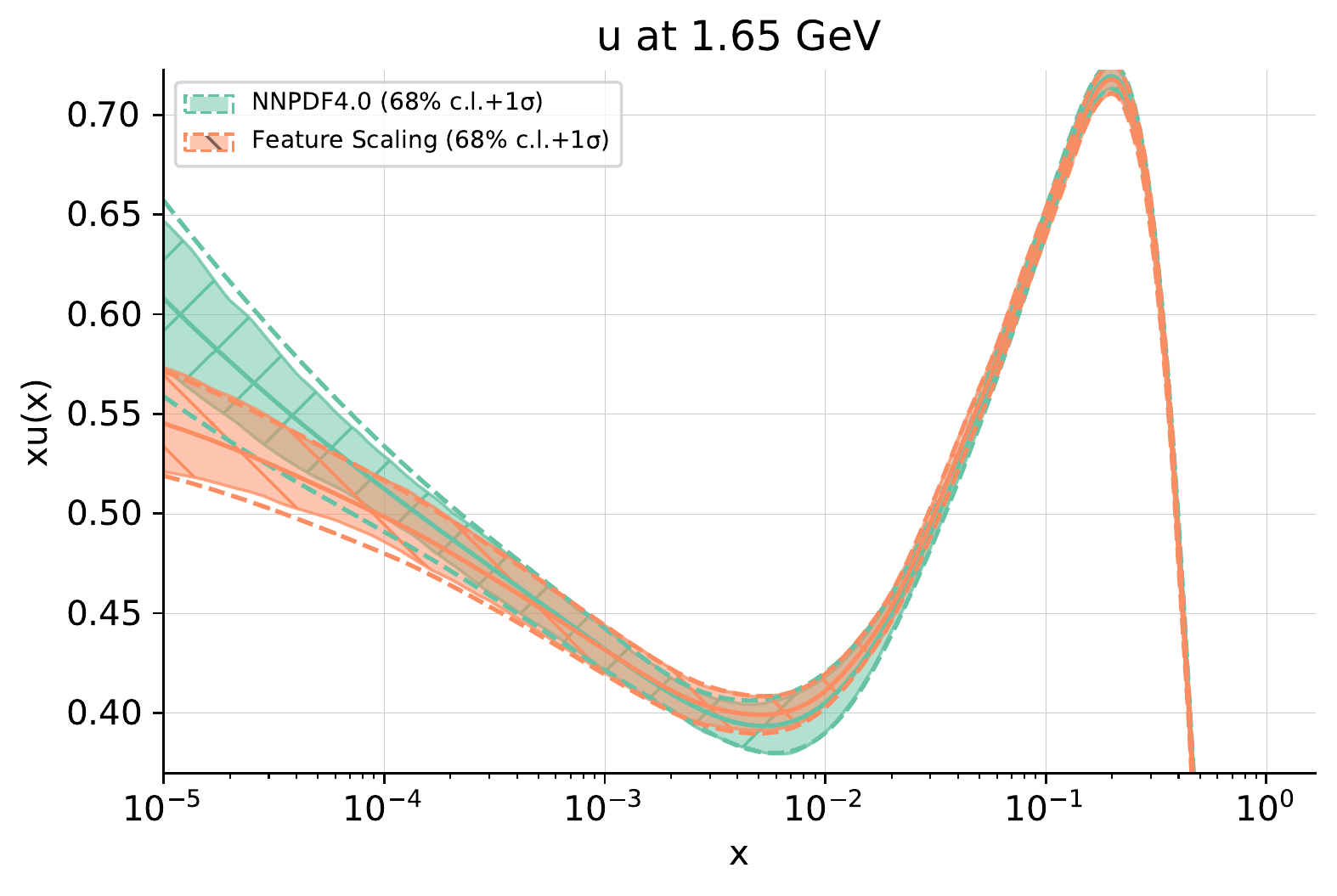}
    \caption{Comparison of the gluon and $u$ quark PDFs between a fit performed with the
        NNPDF4.0 methodology (green), and one with the feature scaling methodology
    (orange).}
    \label{fig:pdf_feature_vs_nnpdf40}
\end{figure}

As can be seen in Fig.~\ref{fig:pdf_feature_vs_nnpdf40}, where we compare the
gluon and $u$ quark PDFs of the NNPDF4.0 fit, to a PDF generated using the
feature scaling methodology, the plots show good agreement between the two PDFs.
While only the two partons are shown, this is representative of all flavors. The
prediction of the feature scaling methodology in the extrapolation region is
validated by performing a future test of the feature scaling methodology as has
been done before for the NNPDF4.0 methodology in Ref.~\cite{Ball:2021leu}. The
results of this future test results shown in Tab.~\ref{tab:future_test_chi2}.
Each column corresponds to a fit perform using all previous datasets
(for instance, the pre-LHC fit includes all the data in pre-HERA as well).
Instead, each row corresponds to the partial dataset used to compute the $\chi^{2}$.
We make a distinction between ${\chi^2}$ inside parentheses with a covariance matrix as defined in
\eqref{eq:cov=exp}, and the ${\chi^2}$ without parentheses corresponding to a
covariance matrix as defined in \eqref{eq:cov=exp+pdf}. Before seeing these results one may wonder whether, because all datasets are sensitive to the same large-$x$ region, the datasets are consistent and
thus the test is trivial.
The answer to this becomes clear by looking at the $\chi^2$ values inside the parentheses which indicate that when the PDF uncertainties are not considered the fit quality is very poor for unseen data.

We can analyze the result starting on the third row corresponding to the NNPDF4.0 dataset.
For the fit that included the entire dataset (third column) it makes virtually
no difference whether or not the PDF uncertainties are taken into account. This
is quite different for the pre-HERA fit (first column): even though the
central PDF is off ($\chi^{2} = 7.23$), once
its uncertainties are considered, the quality of the fit
is comparable to that of NNPDF4.0 with with a $\chi^{2}$ of 1.29
compared to 1.21.

In the second row instead the pre-LHC dataset is considered. Both the NNPDF4.0
and the pre-LHC fit, where the dataset is included, produce a trivially good
$\chi^{2}$ for their fitted data. When we compute the prediction using the
pre-HERA fit instead the number is much worse. Once again, upon considering the
PDF uncertainties, the number is of order one, though still significantly larger
than the corresponding values in the fits with pre-LHC or NNPDF4.0 data. This
suggests that qualitatively good agreement is obtained but stability upon
changes to the dataset can still be improved.
It should be noted that in all cases the methodology used has been hyperoptimized for the full NNPDF4.0 dataset.

If we compare these results as presented in Tab.~\ref{tab:future_test_chi2} for
the feature scaling methodology, to the results for the NNPDF4.0 methodology
shown in Tab.~\ref{tab:future_test_chi2_nnpdf40}, we observe much the same
properties. Indeed, even in cases in which the out-of-sample $\chi^{2}$ of
\eqref{eq:cov=exp} differs greatly between both methodologies, the results are
compatible once the PDF uncertainties are considered. While a strict passing
criterion for the future test, such as a specific threshold $\chi^2$, has never
been defined, it allows us to test whether, when PDF uncertainties are considered,
the agreement to out-of-sample data is of a similar level as that of fitted data
where the PDF uncertainty is not considered. As said, this is indeed the case.

We must note however a deterioration of the results in
Tab.~\ref{tab:future_test_chi2} with respect to those of
Tab.~\ref{tab:future_test_chi2_nnpdf40} which points to a greater dependence on
the considered dataset with the feature scaling methodology.
However, while the NNPDF4.0 set of PDFs has been finely
tuned and highly optimized over many iterations and at high computational
costs, the feature scaling methodology has not been subjected to the same degree of finetunig.
It should also be noted that the NNPDF preprocessing ranges are determined
using the full dataset while the aim of the feature scaling methodology
that can directly accommodate different datasets.
This could also explain why the out-of-sample $\chi^{2}$ of feature scaling
is actually better than that achieved by NNPDF4.0

\begin{table}[ht]
    \resizebox{\columnwidth}{!}{%
        \begin{tabular}{lcccc}
            \toprule
            Dataset  & $N_\mathrm{dat}$ & pre-HERA fit & pre-LHC fit & NNPDF4.0 fit \\
            \midrule
            pre-HERA & 2076 & {    0.87 (0.92)} & {    0.91 (1.03)} & {0.98 (1.08)} \\
            pre-LHC  & 1273 & {\bf 1.35 (5.61)} & {    1.17 (1.27)} & {1.18 (1.20)} \\
            NNPDF4.0 & 1269 & {\bf 1.29 (7.23)} & {\bf 1.22 (4.72)} & {1.21 (1.29)} \\
            \bottomrule
        \end{tabular}
    }%
    \caption{${\chi^2}$ values per datapoint as obtained during a future test of
        the feature scaling methodology. The columns correspond to fits based on a
        given dataset, while the rows correspond to the datasets for which the
        ${\chi^2}$ values are shown. While for the fit the dataset are inclusive
        (i.e., the NNPDF4.0 fit includes also the pre-LHC and pre-HERA datasets) the
        $\chi^{2}$ is computed in an exclusive manner (i.e., the $\chi^2$ as calculated for the NNPDF4.0 dataset only uses ``post-LHC'' data).
        The values in bold represent the performance on datasets that were not part
        of the training. The values inside parentheses correspond to a ${\chi^2}$
        defined with $\sigma$ as defined in \eqref{eq:cov=exp+pdf}, while those
        without parenthesis are defined with only the experimental covariance matrix
        of \eqref{eq:cov=exp}.}
    \label{tab:future_test_chi2}
\end{table}

\begin{table}[ht]
    \resizebox{\columnwidth}{!}{%
        \begin{tabular}{lcccc}
            \toprule
            Dataset  & $N_\mathrm{dat}$ & pre-HERA fit & pre-LHC fit & NNPDF4.0 fit \\
            \midrule
            pre-HERA & 2076 & {    0.87 (0.91)} & {    0.94 (1.01)} & {1.01 (1.06)} \\
            pre-LHC  & 1273 & {\bf 1.22 (26.1)} & {    1.18 (1.21)} & {1.17 (1.20)} \\
            NNPDF4.0 & 1269 & {\bf 1.28 (22.6)} & {\bf 1.28 (2.15)} & {1.23 (1.29)} \\
            \bottomrule
        \end{tabular}
    }%
    \caption{Same as Tab.~\ref{tab:future_test_chi2} for the NNPDF4.0 methodology.}
    \label{tab:future_test_chi2_nnpdf40}
\end{table}

\subsection{Evaluation of large-\texorpdfstring{$x$}{x} extrapolation}
\label{sec:largex}

Upon removing the prefactor, we not only affect the small-$x$
extrapolation region of the PDFs, but also the large-$x$ extrapolation region.
So far no test has been developed to test the validity in this range of the PDF
domain. Using the future test to also test the faithfulness of the predictions
in the large-$x$ region cannot be done in the same way due to the limitations
of the datasets
that do not contain any large-$x$ datapoints (irrespective of how we define
large-$x$ precisely). For example, removing all datasets which contain a point
in $x\gtrsim 0.3$ leaves a set of datasets which do not provide sufficient
constraints on the PDF to perform the future test. Nevertheless, here we will
assess the large-$x$
extrapolation behavior of the PDF produced with feature scaling.

\begin{figure}[ht]
    \includegraphics[width=1\columnwidth]{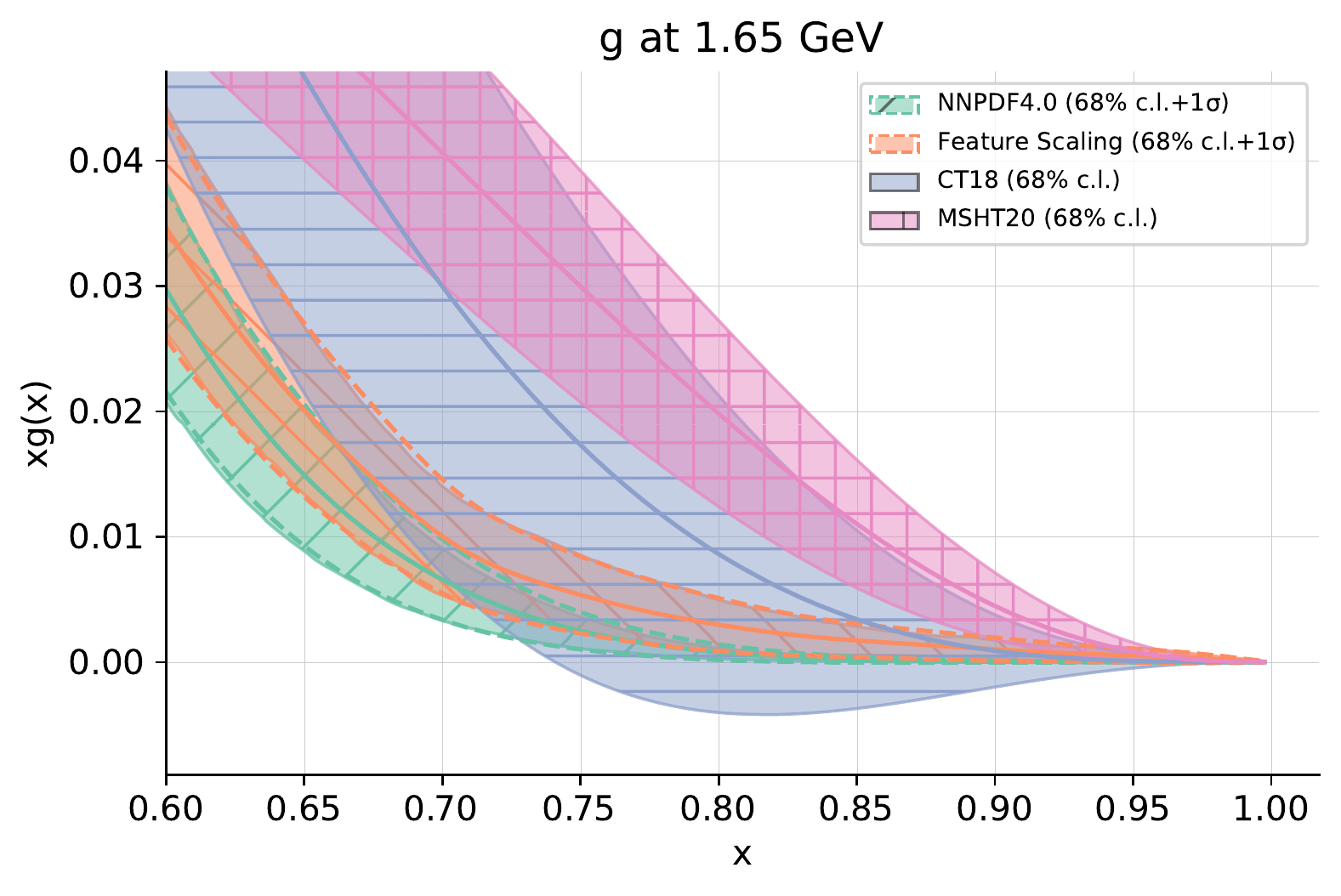}\\
    \includegraphics[width=1\columnwidth]{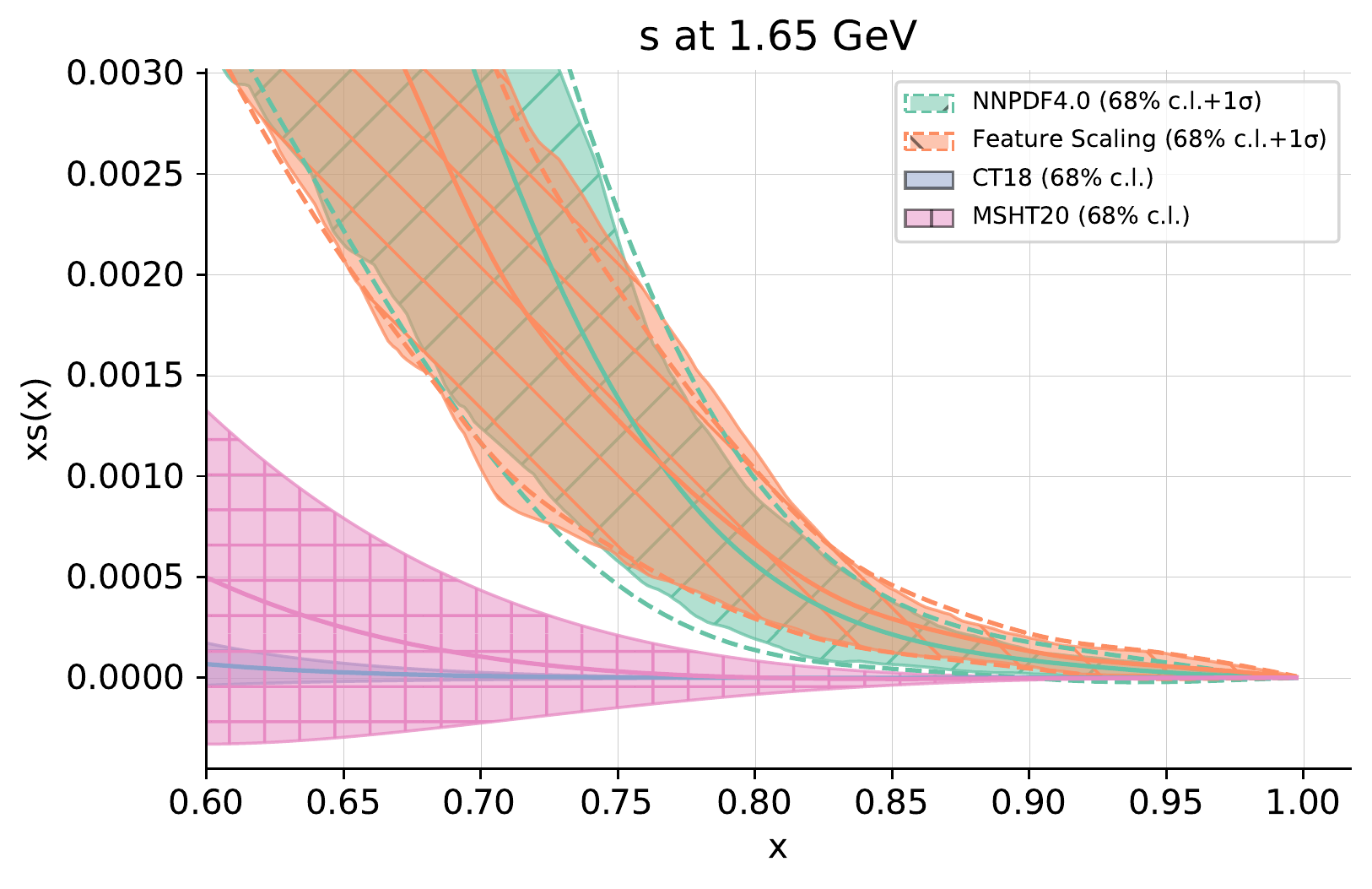}
    \caption{Comparison of the large-$x$ extrapolation regions of the gluon (top)
    and the strange (bottom) PDFs between NNPDF4.0 (green), feature scaling (orange), CT18~\cite{Hou:2019efy} (blue), and MSHT20~\cite{Bailey:2020ooq} (pink).}
    \label{fig:largex_pdfs}
\end{figure}

To do so, let us have a look at the PDFs themselves in this region, and how the
PDFs based on the NNPDF4.0 methodology compare to those that have been produced
with feature scaling. A comparison of the gluon and strange PDF in the domain
$0.6<x<1$ is shown in Fig.~\ref{fig:largex_pdfs}. Note here that there is no
data available for $x>0.75$, meaning that what is shown is mostly
extrapolation region, and these representative examples show a good agreement between the NNPDF4.0 PDF and the feature scaling counterpart. We further want to point out that due to the lack of data in this region different, but all reasonable, parametrization choices can lead to very different results, as can be seen by comparing the NNPDF PDFs to those produced by MSHT or CT.

As a more rigorous check of the large-$x$ extrapolation region one could create
pseudodata based on predictions corresponding to PDFs that have a different
(exponential) behavior in the extrapolation region, e.g. a change of the
$\beta_a$ exponent outside the data region. One can then perform a future test
to this pseudodata, to quantify how well the PDFs generalize in the
extrapolation region. The development of such a test, however, is well beyond
the test of validity we provide here for the feature scaling methodology.
Nevertheless, it can be an interesting check for a future release of PDF sets.

\subsection{Validation of the data region}
\label{sec:closuretest}

Where in Sect.~\ref{sec:futuretest} we performed a future test to validate the
faithfulness of the PDFs in the extrapolation region where the PDFs are not
constrained by data. Here, instead, we will validate the faithfulness of the
PDFs in the data region by performing a closure test as first introduced in
Ref.~\cite{Ball:2014uwa} and extended in the recent NNPDF4.0 paper. Below we
repeat the closure test as performed in Sect.~6.1 of Ref.~\cite{Ball:2021leu} for the
feature scaling methodology, and unless stated otherwise, the same
settings are used.

When fitting experimental data we are subject to complexities in the data such
as inconsistencies between datasets or limitations of the theoretical
calculations. These complexities make it more difficult to assess the
performance of a fitting methodology by analyzing the result of a fit to
experimental data. This realization is what led to the idea of a closure test,
where,
instead of fitting to experimental data, a fit to pseudodata is performed. This
pseudodata is generated by taking a fitted PDF as input, and from that
calculating the observables corresponding to those in the experimental
datasets, thereby creating a dataset with an associated, and known, underlying
PDF. This allows us to test whether our methodology is able to faithfully
reproduce the underlying PDF. To test whether our methodology was successful, a
number of statistical estimators are considered that we will discuss next. For
a detailed motivation of these estimators we refer the reader to
Ref.~\cite{Ball:2021leu}.
As underlying truth we use one non-central replica from a
feature scaling fit.

A first statistical estimator to consider is the $\Delta_ {\chi^2}$
\begin{equation}
    \Delta_ {\chi^2} = {\chi^2}[f^{\rm (cv)}]-{\chi^2}[f^{\rm (ul)}],
\end{equation}
where ${\chi^2}[f^{\rm (cv)}]$ is the loss evaluated for the expectation value
of the fitted model predictions, while ${\chi^2}[f^{\rm (ul)}]$ is the loss
evaluated for the predictions of the PDF used as underlying law. This latter
loss does not vanish, because the pseudodata includes a Gaussian random noise
on top of the central value predictions made using the underlying law. As such,
$ \Delta_ {\chi^2}$ can be understood as an indicator for overfitting or
underfitting: if $\Delta_ {\chi^2}>0$, that indicates underfitting, while
$\Delta_ {\chi^2}<0$ indicates overfitting. For the feature scaling
methodology, the average $\Delta_ {\chi^2}$ as evaluated over observables
corresponding to the full NNPDF4.0 dataset is $\Delta_ {\chi^2}=-0.002$ (compared to $\Delta_ {\chi^2}=-0.009$ for NNPDF4.0), which
is at the per mille level indicating a negligible amount of overfitting.

Now let us estimate the faithfulness of the PDF uncertainty at the level of observables. For this we use the bias over variance ratio
as defined in Eq.~(6.15) of Ref.~\cite{Ball:2021leu}. Here ``bias'' can be
understood as a measure of the fluctuations of the observable values with
respect to the central value prediction of the fitted PDF, while ``variance''
can be understood as the fluctuations of the fitted PDF with respect to its
central value prediction. Thus if the methodology has faithfully reproduced the
uncertainties in the underlying data (bias), this uncertainty should be equal
to the uncertainty in the predictions of the PDFs (variance), and hence the
bias to variance ratio $R_{\rm bv}$ is expected to be one.

To test this, the value of $R_{\rm bv}$ is determined for out-of-sample data.
Specifically, we fit the PDFs to the NNPDF3.1-like dataset as defined in
Ref.~\cite{Ball:2021leu}, and then evaluate the value of $R_{\rm bv}$ for the
data that is part of the NNPDF4.0 dataset but has not already been included in
the NNPDF3.1-like dataset. This allows us to test how well the predication made
using a PDF fitted with a given methodology generalizes to unseen data. The
value of the bias to variance ratio found for the new methodology is $R_{\rm
bv}=1.03\pm0.04$ (compared to $R_{\rm bv}=1.03\pm0.05$ for NNPDF4.0), where again the uncertainty corresponds to a $1\sigma$
bootstrap error, meaning the agreement to the expected value of $R_{\rm bv}=1$
is at the $1\sigma$ level.

To estimate the faithfulness of the PDF uncertainty at the level of the PDF we
calculate a quantile estimator in PDF space $\xi^{(\rm pdf)}_{1\sigma}$. This
quantity corresponds to the number of fits for which the $1\sigma$ uncertainty
band covers the PDF used as underlying law. This is determined for fits
performed to pseudodata covering the full NNPDF4.0 dataset. The result is
$\xi^{(\rm pdf)}_{1\sigma}=0.70\pm0.02$ (compared to $\xi^{(\rm
pdf)}_{1\sigma}=0.71\pm0.02$ for NNPDF4.0), where the uncertainty is a $1\sigma$
uncertainty determined through bootstrapping~\cite{Efron:1979bxm,Efron:1986hys}.
Thus the observed $\xi^{(\rm pdf)}_{1\sigma}$ value is in agreement with the
expected value of 0.68 within $1\sigma$.

An analogous estimator can be calculated for the theory predictions in data
space as opposed to PDF space, providing a generalization to quantile statics of
the bias of variance ratio $R_{\rm bv}$. Similar to the bias over variance
ratio, also for this estimator the values are calculated on out-of-sample data,
where the PDFs have been determined using NNPDF3.1-like data. The expected value
of this quantile estimator depends on the bias over variance ratio is
$\mathrm{erf} (R_{\rm bv}/\sqrt{2}) = 0.67 \pm 0.02$ (compared to $\mathrm{erf}
(R_{\rm bv}/\sqrt{2}) = 0.67 \pm 0.03$ for NNPDF4.0), which is in agreement with
the calculated value of $\xi^{\rm (exp)}_{1\sigma} = 0.69 \pm 0.02$ (and
$\xi^{\rm (exp)}_{1\sigma} = 0.68 \pm 0.02$ for NNPDF4.0).

\section{Conclusions}

In this paper we propose a series of modifications to the PDF parametrization
through the treatment of the input data with a model-agnostic approach.
The result is a simplified fitting procedure which opens the door for further
automatization.

The implications of this new approach have been studied in the context of the
NNPDF fitting framework. We tested the resulting methodology by its own merit
and compared it to the latest release of NNPDF. We found agreement not only
within the data region but also in the extrapolation region (where the choice of
the prefactor can potentially have an effect) which further confirms the
resilience of the NNPDF methodology upon parametrization changes.

This opens up new possibilities for further development of the methodology.
In particular, the hyperoptimization procedure for PDFs
is very sensitive to statistical fluctuations.
Removing randomly chosen exponents eliminates statistical noise and greatly reduce
the number of iterations need to get the best model.

Furthermore, while the fixed scaling of the input data or the choice
of the preprocessing ranges to be sampled need to be reassessed
when the dataset changes considerably
(especially if the extrapolation regions change as a consequence)
the proposed \emph{feature scaling methodology} can accommodate
data changes automatically, becoming more robust as the amount of data increases.

\section*{Acknowledgements}
The authors thank Stefano Forte, Juan Rojo and Richard Ball
for a critical reading of the manuscript and several useful comments,
and Christopher Schwan for providing interpolation grids for cross sections
for the validation of the resulting PDF fits at various stage of this work.
This project is supported by the European Research Council under the European
Unions Horizon 2020 research and innovation Programme grant agreement number
740006.

\FloatBarrier 
\bibliographystyle{epj}
\bibliography{../blbl}

\end{document}